
\input harvmac
\def\pf{{\rm Pf ~}}
\def\np#1#2#3{Nucl. Phys. B{#1} (#2) #3}
\def\pl#1#2#3{Phys. Lett. {#1}B (#2) #3}

\def\prep#1#2#3{Phys. Rep. {#1} (#2) #3}

\Title{hep-th/9402044, RU-94-18}
{\vbox{\centerline{Exact Results on the Space of Vacua of}
\centerline{Four Dimensional SUSY Gauge Theories}}}
\bigskip
\centerline{Nathan Seiberg}
\smallskip
\centerline{\it Department of Physics and Astronomy}
\centerline{\it Rutgers University, Piscataway, NJ 08855-0849}
\bigskip
\baselineskip 18pt
\noindent
We consider four dimensional quantum field theories which have a
continuous manifold of inequivalent exact ground states -- a moduli
space of vacua.  Classically, the singular points on the moduli space
are associated with extra massless particles.  Quantum mechanically
these singularities can be smoothed out.  Alternatively, new massless
states appear there.  These may be the elementary massless particles or
new massless bound states.

\Date{2/24}

\newsec{Introduction}

Supersymmetric quantum field theories are easier to analyze and are more
tractable than non-supersymmetric theories.  The main tool which makes
them simple are the constraints which follow {}from supersymmetry.  In
particular, the holomorphicity of the superpotential when combined with
global symmetries enables one to find many exact results.  For example,
recently it has been shown that both the perturbative non-renormalization
theorems and a new non perturbative generalization of them are simple
consequences of these principles
\ref\nonren{N. Seiberg, \pl{318}{1993}{469}}.
Other exact results about the superpotential which do not follow
trivially {}from the symmetries will be presented in
\ref\ils{K. Intriligator, R. Leigh and N. Seiberg, Rutgers preprint, in
preparation}.

One application of these theories is for dynamical SUSY breaking.  We
will have nothing new to say about it here.  Instead, we'll focus on
another application.  Since some observables in these theories are
exactly calculable, these theories are interesting arenas for the study
of dynamical effects in strongly coupled four dimensional quantum field
theories.  For example, here we will present theories which exhibit
surprising patterns of chiral symmetry breaking and massless bound
states and will argue that some theories have a non trivial critical
behavior associated with massless interacting gluons.

Many of these theories have classical flat directions and hence the
classical theory has a space of inequivalent ground states.  We will
refer to this space as the ``classical moduli space.''  It is singular
at the points where the number of massless fields is increased.  The
degeneracy between these states cannot be lifted perturbatively.  In
some cases non perturbative effects generate a superpotential on the
space which destabilizes these vacua.  In other cases, no superpotential
is generated and the vacuum degeneracy cannot be lifted.  Then it is of
interest to study the quantum moduli space. Of particular interest is
the fate of the singularities on the classical space.

This situation is analogous to two different problems in string theory.
First, classical string theory has moduli spaces which can be studied in
the $\alpha^\prime $ (large radius) expansion.  It is known that the
classical (in the sigma model sense) moduli space is different than the
quantum one.  World sheet instantons modify the space, can change its
topology and can even connect it to a different space.  The second
stringy analogue of our moduli spaces are in the space time
interpretation.  Some of the string moduli spaces might be exactly
stable in the full quantum string theory (for example, if there are
several unbroken supersymmetries in space time).  Then, it is
interesting to know how the classical moduli space is modified quantum
mechanically.  For example, it is known that classically the dilaton
Kahler potential is $K=\log (S+S^\dagger)$.  If the number of space time
supersymmetries is larger than one the vacuum degeneracy associated with
the expectation value of $S$ might not be removed.  What is then the
quantum Kahler potential?  Can the strong coupling region of small
${\rm Re}~ S$ be absent?

We do not have a complete theory of these moduli spaces and their nature
in many cases is not clear to us.  However, we will discuss here three
examples.  In the first, the quantum moduli space is different than the
classical one and the classical singularities are smoothed out.  In the
second example, the quantum space is the same as the classical one but
the physical nature of the singularities is different.  In the classical
theory the singularities corresponds to massless gluons and in the
quantum theory it is associated with massless bound states.  In our
third example the classical and quantum moduli spaces are identical.
Furthermore, both the classical and the quantum theory have massless
gluons and quarks at the singular points.  In the quantum theory these
massless interacting fields correspond to a non trivial four dimensional
critical point.

In section 2 we review the classical field theory of supersymmetric QCD
and present the classical moduli space of the massless quark theory.  In
section 3 we review some known results about the quantum theory.  For
massless quarks the quantum theory with fewer flavors than colors has no
ground state.  When the number of flavors is larger or equal the number
of colors the theory has a quantum moduli space of inequivalent ground
states.  Section 4 discusses the situation for equal numbers of flavors
and colors where the classical singularities on the moduli space are
blown up.  On this space we find points with unusual patterns of chiral
symmetry breaking.  Section 5 discusses the case when the number of
flavors is one plus the number of colors.  The quantum moduli space is
the same as the classical one but the interpretation of the
singularities is different.  At the singular points there are new
massless bound states and some or even all of the chiral symmetry of the
model is unbroken.  Our understanding of the situation with larger
number of flavors is limited.  We present it in section 6.  We speculate
that the singular points might be associated with massless interacting
gluons.  Unfortunately, we can prove that this is the case only for some
range of $N_c$ and $N_f$.

After completing this work we received a paper
\ref\kaplu{V. Kaplunovski and J. Louis, UTTG-94-1,LMU-TPW-94-1}
which partially overlaps with ours.

\newsec{Classical moduli spaces}

We will be studying supersymmetric QCD.  The theory is based on an
$SU(N_c)$ gauge theory with $N_f$ flavors of quarks, $Q^i$ in the $N_c$
representation and $\tilde Q_{\tilde i}$ in the $\bar N_c$
representation ($i, \tilde i =1, \dots, N_f$).  The anomaly free global
symmetry is
\eqn\globsym{SU(N_f)\times SU(N_f) \times U(1)_B \times U(1)_R }
where the $U(1)_R$ charge of $Q$ and $\tilde Q$ is ${N_f-N_c \over N_f}$.
For $N_c=2$ there are $2N_f$ quarks, $Q^i$. The global symmetry is
\eqn\globsymt{SU(2N_f) \times U(1)_R }
and the $U(1)_R$ charge of $Q$ is ${N_f-2\over N_f}$.

An important property of these theories is the existence of classical
flat directions.  For $N_c=2$ the classical flat directions (up to gauge
and global symmetries) are
\eqn\clasflaelt{Q=\pmatrix{a&  &  &  &  \cr
               &a&  &  &  \cr} \quad .}
They can be described by the gauge invariant combinations
\eqn\gaugeinvadeft{V^{ij}=Q^iQ^j \quad .}
The classical moduli space can be described as the space of $V$'s
subject to
\eqn\clasconstt{\epsilon_{i_1, \dots, i_{2N_f}} V^{i_1i_2}
V^{i_3i_4}=0}
(which is meaningful only for $N_f \ge 2$).  This constraint equation
can also be understood as a trivial consequence of the Bose statistics
of the underlying quark superfields.

For non-zero $V$ the gauge symmetry is completely broken and the global
symmetry \globsymt\ is broken to
\eqn\unbrot{SU(2) \times SU(2N_f-2 ) \times U(1)_R}
where this $U(1)_R$ is a linear combination of the original R charge and
a generator in $SU(2N_f)$.  The massless components in $V$ are in the
$(2,2N_f-2)_{N_f-2 \over N_f-1}+(1,1)_0$ representation of the unbroken
global symmetry \unbrot.  One of the scalars in $(1,1)_0$ represents the
inequivalent flat directions labeled by $a$ and all the other scalars
are Goldstone bosons.

Similarly, for $N_c >2$ the flat directions can be labeled by the gauge
invariant combinations (``mesons'' $M$ and ``baryons'' $B,~ \tilde B$)
\eqn\gaugeinvadef{\eqalign{
&M^i_{\tilde i}=Q^i \tilde Q_{\tilde i}\cr
&B_{i_{N_c+1}, i_{N_c+2}, \dots, i_{N_f}}= {1 \over N_c!}
\epsilon_{i_1, \dots, i_{N_f}} Q^{i_1} Q^{i_2}\dots Q^{i_{N_c}} \cr
&\tilde B^{\tilde i_{N_c+1},\tilde  i_{N_c+2}, \dots,\tilde
i_{N_f}}={1 \over N_c!}\epsilon^{\tilde i_1, \dots,\tilde  i_{N_c}}
\tilde Q_{\tilde i_1}\tilde  Q_{\tilde i_2}\dots\tilde  Q_{\tilde
i_{N_c}} \cr }}
whose $U(1)_B\times U(1)_R$ charges are $(0,2{N_f-N_c \over N_f})$,
$(N_c,N_c{N_f-N_c \over N_f})$ and $(- N_c,N_c{N_f-N_c \over N_f})$.
For $N_f <N_c$ the baryons $B$ and $\tilde B$ do not exist and the flat
directions are the space of $M$'s.  For $N_f \ge N_c$ flat directions
are the space of $M$, $B$ and $\tilde B$ subject to the constraints
following {}from Bose statistics of the fundamental quarks. For $N_f=N_c$
there is only one constraint:
\eqn\clasconstn{\det M - B \tilde B =0 \quad .}
For $N_f=N_c+1$ there are three constraints
\eqn\clasconst{\eqalign{
&{1 \over N_c !}\epsilon_{i_1,\dots , i_{N_f}}\epsilon^{\tilde
i_1,\dots, \tilde i_{N_f}} M^{i_1}_{\tilde i_1}\dots M^{i_{N_c}}_{\tilde
i_{N_c}} - B_{i_{N_f}}\tilde B^{\tilde i_{N_f}}= 0 \cr
&B_{i} M^{i}_{ \tilde j} =0 \cr
&M^{j}_{ \tilde i} \tilde  B^{\tilde i}=0  \quad . }}

Various points on the classical moduli spaces exhibit a different
unbroken gauge and global symmetry.  The unbroken global symmetry can
easily be identified by examining the expectation values of the gauge
invariant order parameters \gaugeinvadeft\gaugeinvadef.  The generic
point on the moduli space has an unbroken $SU(N_c-N_f)$ gauge symmetry
which is absent for $N_f \ge N_c-1$.  Special points where $B=\tilde
B=0$ and $M$ has fewer than $N_c-1$ non zero eigenvalues ($V=0$ for
$N_c=2$) have enhanced gauge symmetries.  At these points the classical
moduli spaces is singular.

\newsec{The Quantum theory}

The quantum theory was studied by several groups using different
techniques.  The authors of
\ref\cerne{G. Veneziano and S. Yankielowicz, \pl{113}{1982}{321};
T.R. Taylor, G. Veneziano, and S. Yankielowicz, \np{218}{1983}{493}}
advocated the use of an effective Lagrangian involving the light meson
fields $M$ and the glueball field $S={1 \over 32 \pi^2}W_\alpha^2$ and
imposed the anomalous Ward identities on the superpotential. On the
other hand, reference
\ref\dds{A.C. Davis, M. Dine and N. Seiberg, \pl{125}{1983}{487}}
followed the Wilsonian approach and focused only on the light fields.
Then, the anomalous Ward identities should certainly not be imposed.
Since we are interested here in the moduli space, we should follow the
point of view of \dds\ and keep only the light fields.  In general, a
Wilsonian effective action with at most two space time derivatives must
include all the massless fields.  It may include some of the massive
fields after others have been integrated out.  In this case it
reproduces the correct dynamics of the light fields but might lead to
incorrect answers for the massive ones.

Dynamical calculations in these theories were performed using instanton
methods.  Reference
\ref\ads{I. Affleck, M. Dine, and N. Seiberg, \np{241}{1984}{493};
\np{256}{1985}{557}}
studied these theories along the flat directions and performed the
instanton calculations in the Higgs picture.  The CERN group
\ref\cern{D. Amati, K. Konishi, Y. Meurice, G.C. Rossi and G. Veneziano,
\prep{162}{1988}{169} and references therein}
applied the instanton method of
\ref\itep{V.A. Novikov, M.A. Shifman, A.I. Vainshtein and V.I. Zakharov,
\np{223}{1983}{445}}
and considered the theory near the origin in field space $Q=0$.
For $N_f< N_c -1$ the two approaches lead to qualitatively consistent
answers (reference
\ref\newrus{V.A. Novikov, M.A. Shifman, A.I. Vainshtein and V.I.
Zakharov, \np{260}{1985}{157}}
claims that the details are not consistent).  For $N_f \ge N_c$ the
authors of \cern\ did not agree with the conclusions of \ads.

To control the theory along the flat directions we can add mass terms to
the superpotential $m_i^{\tilde i} Q^i \tilde Q_{\tilde i}$.  Later we
will also add terms of the form $bB+\tilde b \tilde B$ for $N_f \ge N_c$.

For $\det m \not=0$ all the flat directions are lifted and classically
$M=B=\tilde B=0$.  Quantum mechanically the expectation values are
different.  Expectation values of lower components of chiral superfields
are holomorphic in $m$ and must respect selection rules under the global
symmetry \globsym\ of the massless theory \cern.  This leads to
\eqn\expec{\eqalign{
M^{i }_{\tilde i} = &<Q^i \tilde Q_{\tilde i}> \sim \Lambda^{3N_c
-N_f\over N_c} (\det m)^{1 \over N_c}\left({1 \over m}\right)^i_{\tilde
i} \cr
B=& \tilde B = 0
\quad . \cr }}
If we also add $b B + \tilde b \tilde B$ the expectation values of $B$
and $\tilde B$ do not vanish.  The phases {}from the fractional power $1
\over N_c$ correspond to $N_c$ different ground states -- exactly as
predicted by the Witten index.  For $N_c=2$ with the mass term $m_{ij}
Q^iQ^j$ these become
\eqn\expect{V^{i j} = <Q^i  Q^{j}> \sim \Lambda^{6 -N_f\over 2} (\pf
m)^{1 \over 2}\left({1 \over m}\right)^{i j} \quad .}

Explicit calculations \ads\cern\ show that the coefficients of order one
in these relations do not vanish.  Therefore, we can redefine $\Lambda$
such that
\eqn\expecn{\eqalign{
M^{i }_{\tilde i} = &<Q^i \tilde Q_{\tilde i}> = \Lambda^{3N_c -N_f\over
N_c} (\det m)^{1 \over N_c}\left({1 \over m}\right)^{i }_{\tilde i} \cr
V^{i j} = &<Q^i  Q^{j}>= \Lambda^{6 -N_f\over 2} (\pf m)^{1 \over
2}\left({1 \over m}\right)^{i j}  \quad . \cr}}
It is crucial that these expectation values are exact and are not just
approximate.

For $N_f < N_c$ the light fields can be represented by $M$.  Its
expectation value \expecn\ can be obtained {}from the effective Lagrangian
\dds
\eqn\nflncdy{W_{dyn}= (N_c-N_f){ \Lambda^{3N_c-N_f\over N_c - N_f} \over
(\det M)^{1\over N_c - N_f}} + m M \quad .}
Again, we should stress that this expression for the superpotential is
exact \nonren.

For $N_f \ge N_c$ an effective Lagrangian description is more subtle.
It is easy to see that using the constrained light fields or the
elementary quark superfields $Q$ and $\tilde Q$ no invariant
superpotential can be written in the massless theory.  Therefore, the
classical vacuum degeneracy cannot be lifted \dds\ads\ and the quantum
theory has a moduli space of ground states.  This is the point which was
questioned in \cern.

It is of interest to find the quantum moduli space of the massless
theory.  By taking $m$ to zero we should find a point on that space.  It
is enough to diagonalize $m$ (for $N_c=2$ bring $m$ to a block diagonal
form with every block proportional to $\sigma^2$) with eigenvalues
$m_i$.  One way to see that a non-trivial quantum moduli space exists is
to note that the limit $m_i \rightarrow 0$ is not smooth and the limit
of the expectation values depends on the way it is taken.  Below we
analyze this problem for different values of $N_f$.

\newsec{The quantum moduli space for $N_f=N_c$}

It is easy to see using \expecn\ that the classical constraint
\clasconstn\ (or \clasconstt\ for $N_c=2$) is modified quantum
mechanically to
\eqn\quanconst{\eqalign{
&\det M - B \tilde B = \Lambda^{2N_c} \cr
&\pf V = \Lambda^4  \qquad {\rm for} ~ N_c=2 \quad .\cr }}
This modification is due to a one instanton effect.  As mentioned above,
the classical constraint follows {}from Bose statistics of the quark
superfields.  However, such a condition does not necessarily apply
quantum mechanically.  Note that these relations are true for every $m$
and not only in the limit $m \rightarrow 0$.

We therefore conclude that the classical moduli space which was defined
by \clasconstt\clasconstn\ is modified quantum mechanically to the space
defined by \quanconst.  Far {}from the singular points of the classical
moduli space where semiclassical analysis is reliable the quantum space
is very similar to the classical one.  However, the quantum modification
is crucial as it ``blows up'' the singularities; the singular
points $B=\tilde B=0$ with at least two vanishing eigenvalues of $M$
($V=0$ for $N_c=2$) are not on the quantum moduli space.

It is amusing to note that since the quantum moduli space is different
than the classical one, the low energy effective Lagrangian may have
solitons.  Such states cannot be understood as solitons on the classical
moduli space.

Some of the points on the manifold \quanconst\ have an enhanced global
symmetry:

\noindent
1. For $N_c=2$ the points related to
\eqn\nctwomo{V=\Lambda^2 \pmatrix{\sigma^2& \cr
& \sigma^2\cr}}
by the $SU(4)$ symmetry break it to $SP(4)$.  This is the natural guess
for the pattern of chiral symmetry breaking in theories with matter in a
pseudo real representation.  What is somewhat unusual is that the R
symmetry is unbroken.  Therefore we should check the 'tHooft anomaly
conditions associated with it.  The high energy fermions are in $2\times
4_{-1} + 3 \times 1_{1} $ of $SP(4)\times U(1)_R$.  The low energy
fields are the fluctuations of $V$ around the expectation value
\nctwomo\ subject to the constraint \quanconst.  Their fermion
components transform like $5_{-1}$ under the unbroken group.  The
non-trivial anomalies which should be checked are
\eqn\anomcho{\matrix{SP(4)^2U(1)_R & - 2d^{(2)}(4)= -d^{(2)}(5) \cr
 & \cr
U(1)_R^3 & 2\cdot 4\cdot (-1)^3 +3=5\cdot(-1)^3 \cr
 & \cr
U(1)_R & 2\cdot 4 \cdot (-1) +3 = 5 \cdot (-1) \cr}}
where $d^{(2)}(r)$ is the quadratic $SP(4)$ Casimir operators in the $r$
representation.  Note that the anomalies at the macroscopic and
microscopic levels are the same.

\noindent
2. One generalization of these points to arbitrary $N_c$ is
\eqn\ncmo{\eqalign{&B=\tilde B=0\cr
&M^i_{\tilde i} = \Lambda^2 \delta^i_{ \tilde i} \cr}}
corresponding to the breaking of the flavor  $SU(N_f) \times SU(N_f) $
symmetry to the diagonal $SU(N_f)$.  Again, since $U(1)_R$ is unbroken,
we should check the 'tHooft anomaly conditions.  The high energy
fermions are in $N_f\times (N_f)_{1,-1} + N_f\times (\bar N_f)_{-1,-1}
+ (N_f^2 -1) \times 1_{0,1}$ under $SU(N_f) \times U(1)_B \times
U(1)_R$.  The low energy fields are the fluctuations of $M$, $B$ and
$\tilde B$ around the expectation values \ncmo\ subject to the constraint
\quanconst.  Their fermion components transform like $(N_f^2 -1)_{0,-1}
+1_{-N_f,-1}+ 1_{N_f,-1} $ under the unbroken symmetry.  The non-trivial
anomalies are
\eqn\anomcht{\matrix{SU(N_f)^2U(1)_R & -N_f d^{(2)}(N_f)-N_f
d^{(2)}(\bar N_f) = - d^{(2)}(N_f^2-1) \cr
 & \cr
U(1)_R^3 & 2N_f^2 (-1)^3 +(N_f^2-1) = (N_f^2-1)(-1)^3 -2 \cr
 & \cr
U(1)_B^2U(1)_R & - 2N_f^2 \cr
 & \cr
U(1)_R & -2N_f^2 + (N_f^2-1)= -(N_f^2-1) -2 \cr}}
where $d^{(2)}(r)$ is the quadratic $SU(N_f)$ Casimir operators in the
$r$ representation.  Again, the anomalies match.

\noindent
3. Another generalization to arbitrary $N_c$ has
\eqn\ncmoa{\eqalign{&B= - \tilde B= \Lambda^{N_f}\cr
&M^i_{\tilde i} = 0\cr}}
where the $SU(N_f) \times SU(N_f) $ chiral symmetry is unbroken and only
the baryon number symmetry $U(1)_B$ is spontaneously broken.  As before,
at these points the 'tHooft anomaly conditions provide a powerful
consistency check.  The low energy fermions are in the $(N_f, \bar
N_f)_{-1} +(1,1)_{-1} $ representation of $SU(N_f) \times SU(N_f)\times
U(1)_R $.  The relevant anomalies which should be checked are
\eqn\anomch{\matrix{SU(N_f)^3 & N_f d^{(3)}(N_f) \cr
 & \cr
SU(N_f)^2U(1)_R & -N_f d^{(2)}(N_f) \cr
 & \cr
U(1)_R^3 & -N_f^2-1 \cr
 & \cr
U(1)_R & -N_f^2-1 \cr}}
where $d^{(3)}(r)$ is the cubic $SU(N_f)$ Casimir operators in the
$r$ representation.  Again, the anomalies match between the macroscopic
and microscopic levels.

The authors of \cern\ have already noticed some of these anomaly
matching conditions which led them to conjecture that states with these
quantum numbers could be massless.  The novelty of our analysis is the
identification of the quantum moduli space which includes both these
special points and the semiclassical flat directions of \ads.

In order to obtain these results {}from an effective Lagrangian we need to
impose the constraint \quanconst.  One way of doing that is using a
Lagrange multiplier field $X$ with the superpotential
\eqn\ncnfsf{\eqalign{
W=& X \left( \det M - B\tilde B - \Lambda^{2N_f} \right) \cr
W=& X\left( \pf V - \Lambda^4 \right) \qquad {\rm for} ~ N_c=2
\quad . \cr}}

\subsec{Perturbations on the quantum moduli space}

\centerline{\it Example 1:  $W_t=m_{i \tilde i} M^{i \tilde i}$}

If the superpotential \ncnfsf\ is perturbed by mass terms; i.e.\ the
tree level superpotential $W_t=m_{i\tilde i} M^{i \tilde i}$ (or
$m_{ij}V^{ij}$ for $N_c=2$) is added to it, the expectation values
\expec\expect\ are found.  In the special case where $n <N_f$ of the
masses vanish we can integrate out the massive quarks and $X$ in the
effective Lagrangian and find $W_{eff}= (N_c - n) {\Lambda_L^{3N_c-n
\over N_c-n} \over \det ' M }$ where the determinant is only over the
massless modes and the low energy scale $\Lambda_L=\Lambda^{2N_f \over
3N_f -n} (\prod_i m_i)^{1 \over 3 N_f-n}$, exactly as in \nflncdy.

\medskip

\centerline{\it Example 2:  $W_t=bB+\tilde b \tilde B$}

A less trivial application with a non trivial moduli space arises when
the massless theory is perturbed by $bB+\tilde b \tilde B$.  Adding this
to \ncnfsf\ we find
\eqn\withb{\eqalign{&B= \pm i \Lambda^{N_c} \sqrt{\tilde b \over b} \cr
&\tilde B = \pm i \Lambda^{N_c} \sqrt{ b \over \tilde b}  \cr
&\det M (M^{-1}) =0 \cr}}
which means that there is a moduli space of solutions where $M$ is
an arbitrary matrix with at most $N_f -2$ non zero eigenvalues.

Semiclassically, at large field strength, this moduli space can be
understood as follows.  The expectation value $Q=\tilde Q$ with only
$N_f-2$ non zero eigenvalues is a flat direction of the theory with the
superpotential.  It breaks the gauge group to $SU(2)$.  $N_f^2-4$ chiral
superfields acquire masses in the Higgs mechanism and eight more {}from
the superpotential.  The remaining $N_f^2 - 4$ light fields include the
Goldstone bosons of the broken generators in the $SU(N_f) \times
SU(N_f)$ global symmetry and the parameters which label the flat
directions.  Since the unbroken $SU(2)$ gauge theory has no light
quarks, it can easily be integrated out.  Its scale $\Lambda_L$ is
determined by $\Lambda^{2N_f} \sim {Q^{2N_f -4} \over b \tilde b
(Q^{N_f-2})^2} \Lambda_L^6$ where the numerator arises {}from the Higgs
mechanism and the denominator {}from the mass term in the
superpotential.  Since $\Lambda_L$ is independent of $Q$, gluino
condensation in the unbroken gauge group cannot lead to a superpotential
for the light fields.  Furthermore, it is easy to use the symmetries of
the problem and to show that no invariant superpotential can be
generated.  Therefore, the flat directions are not lifted.

The gauge invariant description of this moduli space is in terms of the
matrix $M$ constrained to have at most $N_f-2$ non zero eigenvalues.
This manifold is singular when $M$ has fewer than $N_f-2$ non zero
eigenvalues.  The most singular point is $M=0$.  Classically, these
singularities represent enhanced gauge symmetry at these points.
Quantum mechanically, we expect the gluons of these unbroken gauge
symmetry to confine and not to be massless.  Instead, there might be
other massless fields.  Since our order parameter is $M$, we should find
an effective Lagrangian for $M$.  The symmetries including the
explicitly broken ones as in \nonren\ lead to
\eqn\natgue{W= \sqrt {\tilde b b} \Lambda^{N_f} h\left( {\det M \over
\Lambda^{2N_f} } \right) }
for some function $h$.  In the next example we will integrate out $B$
and $\tilde B$ more carefully and will show that $h(t)= \pm
2\sqrt{t-1}$.  The two signs correspond to the two values of $B$ and
$\tilde B$ (for a more detailed discussion of such phenomena see \ils).
The supersymmetric ground states which follow {}from this Lagrangian
satisfy $\det M {1 \over M}=0$ exactly as in the full theory and as
expected in the semiclassical region.  Furthermore, at the singular
points of this manifold, the unbroken global symmetry is enhanced and
more components of $M$ become massless.  They join the other massless
fields to representations of the unbroken symmetry.  In particular, for
$M=0$ the global $SU(N_f) \times SU(N_f)$ is unbroken and all the
components of $M$ are massless.

\medskip

\centerline{\it Example 3:  $W_t=m_i^{\tilde i}M_{\tilde i}^i
+ bB + \tilde b \tilde B $}

We now combine the previous two examples and consider the tree level
superpotential $W_t=m_i^{\tilde i}M_{\tilde i}^i + bB + \tilde b \tilde
B $.  For simplicity, we present the answers only for $N_c=N_f=3$.
Classically, there is one vacuum at the origin $M=B=\tilde B=0$ where
the gauge group is unbroken and another ground state with $M={\det m
\over b\tilde b} {1 \over m}$, $B=-{\det m \over b^2 \tilde b}$, $\tilde
B=-{\det m \over \tilde b^2 b}$ where the gauge group is completely
broken.  The expectation values in the quantum theory are determined by
\eqn\wthreethree{W= X \left(\det M - B\tilde B - \Lambda^6 \right) +
m_i^{\tilde i}M_{\tilde i}^i + bB + \tilde b \tilde B \quad. }
It is easy to see that the expectation values are
\eqn\finalsob{\eqalign{
B&= - {\Lambda^6 \tilde b  \over \det m} y^2 \cr
\tilde B&=- {\Lambda^6 b  \over \det m} y^2  \cr
M &= \Lambda^3 y {1 \over m} \cr}}
where $y$ satisfies
\eqn\eqforz{b \tilde b y^4 - {\det m \over \Lambda^3} y^3 + \left( {\det
m \over \Lambda^3} \right)^2 =0 \quad .}
This equation has four solutions. For small $\det m \over \Lambda^3$
they are
\eqn\smalm{y_{1234}=\xi \left({\det m \over \Lambda^3}\right)^{1\over 2}
(b \tilde b)^{-{1\over 4}} + {\det m \over 4 \Lambda^3 b \tilde b} +
\dots}
with $\xi^4=-1$ and for small $b \tilde b$
\eqn\solz{\eqalign{
y_{123}&= \omega \left( {\det m\over \Lambda ^3 }\right)^{1 \over 3} +
{1\over 3}
\omega^2 \left( {\det m\over \Lambda ^3 }\right)^{-{1 \over 3}} b \tilde
b + \dots \cr
y_4&=  {\det m\over b \tilde b \Lambda ^3} +\dots \cr} }
with $\omega^3=1$.

This example demonstrates that not only can there be a continuous
manifold of inequivalent ground states there can also be inequivalent
discrete vacua.  Semiclassically we found two different ground states.
They are most easily related to the situation for small $b \tilde b$.
For $b \tilde b=0$ the model has only the first three states.  The
anomaly free $Z_6$ R symmetry is spontaneously broken by gluino
condensation to $Z_2$ and the three ground states are related by the
symmetry.  These three states correspond to the unique state we found
semiclassically near the origin.  For small $b \tilde b$ there are four
inequivalent ground states.  The first three solutions near the origin
are no longer related by symmetry.  The fourth one which was also
observed semiclassically is at large field strength.

We can now use the equations of motion of all the fields to integrate
out $B$, $\tilde B$ and $X$ and to find an effective superpotential for
$M$ only and thus determine the function $h(t)$ in \natgue. Generalizing
to arbitrary $N_f=N_c$, it is straightforward to find the equations of
motion and to show that they are reproduced by $W_{eff}=\pm 2 \sqrt{b
\tilde b} \Lambda^{N_f} \sqrt{{\det M \over \Lambda^{2N_f}} -1} + mM $
and therefore, $h(t)= \pm 2 \sqrt{t-1}$.

This example allows us to get new insight about a method of instanton
calculations which was used in \cern\ following \itep.  There, one
examines suitably chosen correlation functions which are saturated by
instantons and can be proven to be independent of the positions of the
operators.  In this example, one expects (we did not calculate it
explicitly)
\eqn\ruscern{\eqalign{
&<B(x_1) \tilde B(x_2)>=0 \cr
& {1\over 6} \epsilon_{i_1 i_2 i_3} \epsilon^{\tilde i_i \tilde i_2
\tilde i_3} < M_{i_1}^{\tilde i_1}(x_1) M_{i_2}^{\tilde i_2}(x_2)
M_{i_3}^{\tilde i_3}(x_3)> = \Lambda^6 \quad . \cr}}
For $b \tilde b=0$ these correlation functions should be interpreted as
an average over the three ground states.  Since they are related by a
symmetry, they all contribute with equal weights.  However, when $b
\tilde b$ is non zero the four ground states contribute with unequal
weights (e.g. for small $b \tilde b$ the fourth one does not contribute
at all).  This is intuitively obvious because there is no symmetry
relating these ground states and the fourth one is very far in field
space and hence has small overlap with the state in which the
expectation values \ruscern\ are computed.  Therefore, without
understanding the nature of the various ground states, it is complicated
to use \ruscern\ to extract the vacuum expectation values in all vacua.

\newsec{The quantum moduli space for $N_f=N_c+1$}

As for $N_f=N_c$, the expectation values \expecn\ do not satisfy
the classical constraints \clasconstt\clasconst
\eqn\quanconstp{\eqalign{
&\det M \left({ 1\over M} \right) _i^{\tilde i} - B_{i}\tilde B^{\tilde
i}= \Lambda^{2N_c-1} m_i^{\tilde i} \cr
& \pf V \left({ 1\over V} \right)_{ij}=\Lambda^3 m_{ij} \quad . \cr}}
However, unlike the $N_f=N_c$ case the classical constraints are
satisfied in the $m_i \rightarrow 0$ limit.  Therefore, the quantum
moduli space of the massless theory is the same as for the classical
theory.  The only exception is at the singular points where different
light fields might be present.  Note that for $m_i \not= 0$ all values
of $M$ and $V$ (and with $bB+ \tilde b \tilde B$ also of $B$ and $\tilde
B$) can be found. Again, this is unlike the $N_f=N_c$ case.  This
suggests that a complete description for non zero $m$ needs all the
fields $M$, $B$ and $\tilde B$ (and all the components of $V$ for
$N_c=2$) and not only the constrained ones.

We now discuss the behavior of the massless theory at the singular
points $V=0$ and its higher $N_c$ analogs like $M=B=\tilde B=0$.  Since
the expectation values of all our order parameters vanish there it seems
like the full chiral symmetry should be unbroken there.  Since we are
planning to use all the components of $M$, $B$ and $\tilde B$ (all the
components of $V$ for $N_c=2$), it is natural to expect that all of them
are massless there.

For $N_c>2$ the massless quarks in the microscopic theory transform
under the global symmetry \globsym\ as $N_c \times (N_f,1)_{1, {1 \over
N_f}}+N_c \times (1, \bar N_f)_{-1, {1 \over N_f}}$ and there are also
$N_c^2-1$ gauge fields.  The fields $M$, $B$ and $\tilde B$ transform
like $(N_f, \bar N_f)_{0, {2 \over N_f}}$, $(\bar N_f, 1)_{N_f-1, {N_f
-1 \over N_f}}$ and $(1, N_f)_{-N_f+1, {N_f -1 \over N_f}}$
respectively.  As a first test we should check the 'tHooft anomaly
equations.  The non trivial ones are
\eqn\anomchtt{\matrix{SU(N_f)^3 & (N_f-1) d^{(3)}(N_f) \cr
 & \cr
SU(N_f)^2U(1)_R & -{(N_f -1)^2 \over N_f} d^{(2)}(N_f) \cr
 & \cr
U(1)_R^3 & -N_f^2 +6N_f -12 +{8 \over N_f} - {2 \over N_f^2} \cr
 & \cr
U(1)_B^2U(1)_R & - 2(N_f-1)^2 \cr
 & \cr
SU(N_f)^2U(1)_B & (N_f -1) d^{(2)}(N_f) \cr
 & \cr
U(1)_R^2U(1)_B & 0 \cr
 & \cr
U(1)_R &  -N_f^2+2N_f -2 \cr}}
and they match between the low energy and the high energy spectra.
Reference \cern\ noticed this anomaly matching and conjectured the
existence of a ground state with these massless particles.

For $N_c=2$, the fundamental quarks are in $6_{1 \over 3}$ and the field
$V$ is in $15_{2\over 3}$ of $SU(6) \times U(1)_R$.  The 'tHooft anomaly
conditions are satisfied:
\eqn\anomchttt{\matrix{SU(6)^3 & 2 d^{(3)}(6)= d^{(3)}(15) \cr
 & \cr
SU(6)^2U(1)_R & 2(-{2 \over 3}) d^{(2)}(6)= -{1 \over 3}d^{(2)}(15) \cr
 & \cr
U(1)_R^3 & 12 (-{2 \over 3})^3 + 3 = 15(-{1 \over 3})^3 \cr
 & \cr
U(1)_R & 12 (-{2 \over 3}) + 3 = 15(-{1 \over 3}) \quad . \cr}}

Given that there are more massless fields at the origin, we should be
able to find a low energy effective Lagrangian which describes all these
fields.  There is a unique superpotential
\eqn\lowsupnfo{\eqalign{
W_{eff} &= {1 \over \Lambda^{2N_f -3}} \left( B_i
M^i_{\tilde i} \tilde  B^{\tilde i} - \det M \right) \cr
W_{eff} &= - {1 \over \Lambda^3} \pf V
\cr}}
which satisfies the following properties:

\noindent
1. It is invariant under all the symmetries in of the problem including
the $U(1)_R$ symmetry.

\noindent
2. Its flat directions are exactly as in the microscopic theory.  The
classical constraint equations \clasconstt\clasconst\ which are not
modified quantum mechanically arise here as the equations of motion
{}from \lowsupnfo.

\noindent
3. At the origin all the fields are massless.  However, away {}from the
origin only some of the fields remain massless in a way consistent with
the semiclassical treatment.

\noindent
4. By adding $m_i^{\tilde i} M^i_{\tilde i} + b^i B_i + \tilde b_{\tilde
i} \tilde B^{\tilde i}$ and solving for the fields we recover all the
results for $N_f <N_c +1$.  In particular, we can add masses to some
of the fields, integrate them out and find the low energy Lagrangian for
fewer flavors.

This theory is similar to the second example in section 4.  There we
also found a singular moduli space where the singularity was associated
with new massless fields.  As in that example, the superpotential leads
to masses to some of the fields away {}from the origin and its equations
of motion are the defining equations of the moduli space.

\newsec{The quantum moduli space for $N_f \ge N_c+2$}

It is not easy to extend the previous descriptions for $N_f \ge N_c+2$.
As for $N_f = N_c+1$, all values of $V$, $M$, $B$ and $\tilde B$ can be
obtained and they should all be considered independent fields.  By
examining the massless limit of the expectation values we see that the
classical constraints are satisfied quantum mechanically.  Therefore, as
for $N_f = N_c+1$, the quantum moduli space is the same as the classical
one.  However, unlike $N_f = N_c+1$, the singularities cannot be
associated with unbroken global symmetry and massless $V$, $M$, $B$ and
$\tilde B$ fields.  This can be seen in several ways.

\noindent
1.The 't Hooft anomaly conditions are not satisfied there.

\noindent
2. An effective Lagrangian description depending only on our mesons and
baryons is singular.  Consider for simplicity the case of $N_c=2$.
There is a unique invariant superpotential for $V$
\eqn\uninv{W= {2- N_f\over \Lambda^{6-N_f \over N_f-2}} {\pf V}^{1 \over
N_f -2} \quad . }
Although it leads to the correct expectation values of $V$ in the
massive theory it is singular at $V=0$.  We could try, following \cern\
to add the field $S$.  Even without imposing the anomalous Ward
identities the symmetries lead to $W= S f \left({\pf V \over S^{N_f-2}
\Lambda^{6-N_f}} \right) $ which is always singular at $V=S=0$.

We conclude that a complete gauge invariant description near the origin
needs more fields.  We could not find a simple set of fields which could
resolve the singularity.  Perhaps the Kahler potential is singular there
and the origin is infinitely far {}from every points on the moduli space.

An alternative is that at least some of the elementary colored fields
are massless at the origin.  One possibility is that the gauge group
breaks to an abelian subgroup there.  A more interesting possibility is
that the spectrum at the origin is identical to that in the classical
theory.  Clearly, this is the simplest solution of the 'tHooft
equations...

Massless quarks and gluons are possible only if the theory is scale
invariant.  We cannot show that such a scale invariant theory exists for
every $N_f \ge N_c+2$.  However, it is easy to establish it at least for
some range of $N_f$ and $N_c$.

The two loop beta function of this theory is
\ref\twoloop{D.R.T. Jones, \np{75}{1974}{531}; R. Barbieri, S. Ferrara,
L. Maiani, F. Palumbo and C.Savoy, \pl{115}{1982}{212};  A. Parkes and
P. West, \pl{138}{1984}{199}}
\eqn\betafunt{\beta(g) = - {g^3 \over 16 \pi^2} (3N_c - N_f) + {g^5
\over 128 \pi^4}  (2 N_c N_f-3N_c^2  - {N_f \over N_c}) + \CO( g^7) }
The authors of
\ref\rusano{V. Novikov, M. Shifman, A. Vainshtein, M. Voloshin and V.
Zakharov, \np{223}{1983}{394}; M.A. Shifman and A.I Vainshtein,
\np{227}{1986}{456}}
argued that the exact beta function satisfies
\eqn\betafun{\eqalign{
\beta(g)&= - {g^3 \over 16 \pi^2} {3N_c - N_f + N_f \gamma(g) \over 1
- N_c {g^2 \over 8 \pi^2}} \cr
\gamma(g) & = -{g^2 \over 8 \pi^2}{N_c^2 -1 \over N_c} + \CO( g^4) \cr}}
where $\gamma(g)$ is the anomalous dimension of the mass (equation
\betafun\ is consistent with \betafunt).

Since there are values of $N_f$ and $N_c$ where the one loop beta
function is negative but the two loop contribution is positive, there
might be a non trivial critical point
\ref\bankszaks{T. Banks and A. Zaks, \np{196}{1982}{189}}.
Several people noticed that by taking $N_c$ and $N_f$ to infinity
holding $N_c g^2$ and ${N_f \over N_c}=3 - \epsilon$ fixed, one can
establish the existence of a critical point at $N_cg_*^2={8 \pi^2 \over
3} \epsilon + \CO(\epsilon^2) $.  Therefore, at least for large $N_c$
and $\epsilon= 3-{N_f \over N_c} \ll 1$ and perhaps even for every $N_f
\ge N_c +2$ there are massless interacting gluons and quarks at the
origin.

We would like to make a few comments about these scale invariant
theories:

\noindent
1. At the origin the operators have anomalous dimensions.  For example,
equation \betafun\ shows that $\gamma(g_*)= 1 -{3N_c \over N_f }$ (in
deriving this result we only need the numerator of \betafun\ which is
better motivated than the full expression).

\noindent
2. $\Lambda$ in the formulas is the scale at which the theory crosses
over from the UV to the IR critical behavior.  Using $\Lambda$ we can
relate the UV expression for the mass to the IR expression which has an
anomalous dimensions.

\noindent
3.  By adding mass terms for the quarks one can flow between these
critical points by reducing $N_f$.

\centerline{{\bf Acknowledgements}}

It is a pleasure to thank T. Banks, K. Intriligator, R. Leigh, G. Moore,
S.  Shenker and E. Witten for many useful discussions.  This work was
supported in part by DOE grant \#DE-FG05-90ER40559.

\listrefs
\bye